\documentclass[aps,onecolumn,showkey,square,numbers,amsthm,amssymb,amsmath,nobibnotes,nofootinbib,10pt]{revtex4}

\usepackage{mathtools}
\usepackage{amsthm}
\usepackage{dsfont}
\usepackage{booktabs}
\usepackage{amstext}
\usepackage{esint}
\usepackage{bm,bbm}  
\usepackage{eucal}
\usepackage{verbatim}
\usepackage{booktabs,array}
\usepackage{times}
\usepackage{graphicx,color}
\usepackage{tabularx}
\usepackage{dsfont}
\usepackage{titletoc} 
\usepackage{microtype}
\usepackage[colorlinks=true,linkcolor=black,citecolor=black,urlcolor=blue]{hyperref}
\usepackage{tikz}
\usetikzlibrary{decorations.markings}
\usepackage{tkz-graph}
\usepackage[caption=false]{subfig}
\usepackage{blkarray}
\usepackage{hhline}

\usepackage{boxedminipage,changebar}

\theoremstyle{plain}
\newtheorem*{theorem}{Theorem}

\theoremstyle{remark}
\newtheorem*{rmk}{Remark}

\newcommand{\barr}{\begin{eqnarray}}
\newcommand{\earr}{\end{eqnarray}}
\newcommand{\be}{\begin{equation}}
\newcommand{\ee}{\end{equation}}

\def\Ord{\mathcal{O}}
\newcommand{\de}{\mathrm{d}}

\newcommand{\abs}[1]{\left| #1 \right|} 
 
\let\baraccent=\= 
\renewcommand{\=}[1]{\stackrel{#1}{=}} 

\newcommand{\numberset}{\mathbb}
\newcommand{\N}{\numberset{N}}

\newcommand{\R}{\numberset{R}}
\newcommand{\C}{\numberset{C}}

\newcommand{\avg}[1]{\left< #1 \right>} 
\newcommand{\tr}{\mathrm{Tr}}
\newcommand{\Tr}{\mathrm{Tr}}

\newcommand{\kk}{\kappa}
\newcommand{\EE}{\mathcal{E}}

\newcommand{\G}{\mathcal{G}}
\newcommand{\W}{\mathcal{W}}
\newcommand{\J}{\mathcal{J}}
\newcommand{\supp}{\mathrm{supp}\,}

\newcommand{\Cov}{\mathrm{Cov}}
\newcommand{\NN}{\mathcal{N}}

\newcommand{\XX}{\mathcal{X}}

\newcommand{\Res}{\mathrm{Res}}

\DeclarePairedDelimiter\floor{\lfloor}{\rfloor}

\begin{document}

\title{A unified fluctuation formula for one-cut $\beta$-ensembles of random matrices}

\author{Fabio Deelan Cunden$^{1,2}$, Francesco Mezzadri$^{1}$, Pierpaolo Vivo$^{3}$}
\address{$1.$ School of Mathematics, University of Bristol, University Walk, Bristol BS8 1TW, England\\
$2.$ Istituto Nazionale di Fisica Nucleare (INFN), Sezione di Bari, I-70126 Bari, Italy\\
$3$. King's College London, Department of Mathematics, Strand, London WC2R 2LS, United Kingdom}

\begin{abstract}
Using a Coulomb gas approach, we compute the generating function of the covariances of power traces for one-cut $\beta$-ensembles of random matrices in the limit of large matrix size. This formula depends only on the support of the spectral density, and is therefore universal for a large class of models.  This allows us to derive a closed-form expression for the limiting covariances of an arbitrary one-cut $\beta$-ensemble. As particular cases of  {the main result} we consider the classical $\beta$-Gaussian, $\beta$-Wishart and $\beta$-Jacobi ensembles, for which we derive previously available results as well as new ones  within a unified simple framework.  We also discuss the connections between the problem of trace fluctuations for the Gaussian Unitary Ensemble and the enumeration of planar maps.
\end{abstract}

\maketitle

\section{Introduction} 
Many textbooks on random matrices~\cite{Tao,Guionnet,Bai} begin with the derivation of Wigner's semicircle law for the Gaussian Unitary Ensemble (\textsf{GUE}),~\footnote{For convenience and consistence with the formalism used for the problem of the enumeration of maps (see Appendix A), throughout this paper the \textsf{GUE} is defined by complex Hermitian $N \times N$ matrices $\G_N$ whose diagonal and off-diagonal elements are normal random variables with the same variance, i.e. $\G_{ij}=\frac{1}{\sqrt{N}}\,x_{ij}\ ,\text{with}\,\, x_{ij}\sim\NN_{\C}(0,1)\,\,\text{for}\,\,i<j\ ,\,\text{and} \,\, x_{ii}\sim\NN_{\R}(0,1)\,\,\text{for all}\,\,i$.} which is arguably the best known result in Random Matrix Theory (\textsf{RMT}). One of the most common proofs  consists in introducing the moments $G_{N,\kk}=N^{-1}\tr\G_N^{\kk}$, where $\G_N$ denotes an $N \times N$ \textsf{GUE} matrix and $\kk$ is an integer, and in applying  Wick's formula to evaluate the averages $\avg{G_{N,\kk}}$.  In order to compute these expectation values, certain diagrams are associated to various contributions of Wick's expansion; then, in the large $N$ limit, only certain diagrams (the planar ones) survive, which can be counted explicitly.  Eventually,  one discovers that the limits $\lim_{N\to\infty}\avg{G_{N,\kk}}$ are non-zero only for $\kk$ even and given by the \emph{Catalan numbers}.   This computation implies that the empirical  density $\rho_N(x)=N^{-1}\sum_i\delta(x-\lambda_i)$ of the eigenvalues $\{\lambda_i\}$ converges in expectation to Wigner's law $\rho(x)=(2\pi)^{-1}\sqrt{4-x^2}$. The simplicity of this result {might lead to} believe that the very same approach (known as the \emph{method of moments}) should be easily extended to higher cumulants of $G_{N,\kk}$. However, it turns out that this problem is highly nontrivial, with very few results available in the literature~\cite{Chekhov06,Guionnet13,Dumitriu06,Dumitriu12,Mingo04,Mingo06,Mingo07,Redelmeir}. To date, almost 60 years after Wigner's original paper~\cite{Wigner57}, a self-contained explicit formula for the higher cumulants $C_v(G_{N,\kk_1},\dots,G_{N,\kk_v})$  for generic $v$ and $\{\kk_i\}$ is still lacking~\footnote{$C_v(\xi_1,\dots,\xi_v)$ denotes the cumulant of order $v$ of the (not necessarily distinct) random variables $\xi_1,\dots,\xi_v$.  For instance $C_2(\xi_1,\xi_2)=\avg{\xi_1\xi_2}-\avg{\xi_1}\avg{\xi_2}$.}. The situation is even worse for general invariant ensembles, where the method of moments is not applicable.

In this work we consider random self-adjoint matrices $\XX_N$ 
whose eigenvalues $\{\lambda_k\}\in\R$ have the canonical distribution of a 2D Coulomb gas at inverse temperature $\beta>0$, namely
\be
\de\mathbb{P}_{N,\beta}(\left\{\lambda_k\right\})=  \frac{1}{\mathcal{Z}_{N,\beta}} e^{-\beta \left(-\frac{1}{2}\sum_{i\neq j}{\ln{|\lambda_i-\lambda_j|}}+N\sum_{i}{V(\lambda_i)}\right)}\prod_{i=1}^{N}\de\lambda_i. \label{eq:cov_jpdf} 
\ee
Here the single particle potential $V(x)$ is assumed to be such that in the thermodynamic (large $N$) limit the Coulomb gas reaches a stable equilibrium distribution supported on a single bounded interval $[a,b]$ of the real line.
This class of \emph{one-cut ensembles}\footnote{The Cauchy-Stieltjes transform of the gas distribution has one connected cut on the real line of the complex plane, hence the name \emph{one-cut}.} includes the classical Gaussian, Wishart and Jacobi ensembles of \textsf{RMT}. For $\beta=1,2$ and $4$,~\eqref{eq:cov_jpdf} is the joint law of the eigenvalues of real symmetric, complex Hermitian or quaternion self-dual invariant matrices, respectively. General ensembles parametrized by non-integer values of $\beta>0$ can be realized from sparse real random matrices, as shown by Dumitriu and Edelman~\cite{Dumitriu02} and Killip and Nenciu~\cite{KN04}. Moreover, in view of certain applications to physics (see, e.g.,~\cite{Forrester}), it is worth studying the general joint probability density function~\eqref{eq:cov_jpdf} independently of a concrete matrix model realization.

In this paper we consider the moments $X_{N,\kk}=N^{-1}\Tr\XX_N^{\kk}$ for matrices $\XX_N$ with joint probability law of the eigenvalues given by~\eqref{eq:cov_jpdf}, and we derive self-contained formulae for their covariances in the limit $N\to\infty$.  More precisely we study the quantities
\begin{equation}
\lim_{N\to\infty}N^2\Cov(X_{N,\kk},X_{N,\ell})= \frac{1}{\beta}\alpha_{\kk,\ell}.\label{eq:first}
\end{equation}
The $1/\beta$-dependence and the $N^{-2}$ scaling of such covariances are customary in the fluctuations for 2D Coulomb gases~\cite{Beenakker93}. {The main result of the paper} is a \emph{universal} formula for the generating function of~\eqref{eq:first}; it is universal in the sense that it depends only on the support of the equilibrium density, but not on the potential $V(x)$.   This is a direct consequence of the macroscopic universality of the smoothed two-point function in \textsf{RMT}.  
Therefore, we can always write the covariances~\eqref{eq:first} in terms of a reference one-cut ensemble --- the natural choice is the \textsf{GUE}.  This allows us to prove a self-contained and manageable formula for the limiting covariances of an arbitrary one-cut ensemble~\eqref{eq:cov_struct2}.  As particular examples, we derive known as well as new results for the Gaussian (Hermite), Wishart (Laguerre) and Jacobi $\beta$-ensembles within a unified simple approach, unveiling the underlying structure. 

We will derive our results in the framework of the Coulomb gas analogy. The advantage is twofold. Firstly, the method is insensitive to the specific value of $\beta>0$, so there is no need to produce separate proofs for each symmetry class. Secondly, it easily applies to matrix models whose underlying diagrammatic is cumbersome or unclear (such as the Jacobi ensemble). Moreover, the method is a good candidate for generalizations to higher order cumulants.

Various methods have been employed to compute cumulants of moments of $\beta$-ensembles. All of them are combinatorial in nature. There are several works on the Gaussian ensembles scattered throughout the literature of quantum field theory in the limit of large internal symmetry group~\cite{Brezin78,Bessis80,Itzykson80,Penner88,Itzykson90}. Some of these results reproduce earlier findings by Tutte~\cite{Tutte62a,Tutte62} for planar enumeration of maps. Johansson~\cite{Johansson98} studied the global fluctuations of Hermitian random matrices and was the first one to prove a central limit theorem for $\beta$-ensemble  with polynomial potentials. More recently, in a series of paper devoted to the second-order freeness~\cite{Mingo04,Mingo06,Mingo07,Redelmeir}, the relation between moments of some random matrix ensembles and non-crossing partitions has been employed to compute covariances to leading order in $N$ on complex and real Gaussian and Wishart matrices. Dumitriu and Edelman~\cite{Dumitriu06} and Dumitriu and Paquette~\cite{Dumitriu12} studied the fluctuations around the equilibrium density in Gaussian, Wishart and Jacobi ensembles at generic Dyson index $\beta$.  Their approach was  based on the realization of these ensembles in terms of sparse matrices. The computation of cumulants of moments for some $\beta$-ensembles is possible in principle by solving a recurrence relation associated to the hierarchy of loop equations~\cite{Chekhov06,Borot11,Guionnet13}.

The plan of this paper is as follows. We start with some preliminaries and a summary of the main ideas of the Coulomb gas approach (Section~\ref{sec:Coulomb}). In Section \ref{sec:main} we present our main result. In Section~\ref{sec:applications} we {analyse} the classical $\beta$-ensembles: the Gaussian, Laguerre and Jacobi $\beta$-ensembles. 
Finally, in Section~\ref{sec:concl} we discuss some open problems.
The paper is complemented by two appendices. In Appendix~\ref{app:review} we review the connections between the problem of moments fluctuations, enumeration of planar maps and non-crossing pair partitions. Appendix~\ref{appB} contains the derivation of a technical result.

\section{2D Coulomb gas and cumulants of linear statistics}
\label{sec:Coulomb}
Here we briefly summarize (without derivations) some aspects of the 2D Coulomb gas approach, which are more directly relevant to our work. The starting point of the method is the analogy with a gas of Coulomb charges pointed out by Wigner and Dyson~\cite{Wigner57,Dyson62}. The joint law $\mathbb{P}_{N,\beta}(\left\{\lambda_k\right\})$ in~\eqref{eq:cov_jpdf} is the Gibbs-Boltzmann measure of a 2D Coulomb gas confined in a interval $\Lambda\subset\R$ at inverse temperature $\beta>0$ (the Dyson index) with a single-particle potential $V(\lambda)$ bounded from below and finite for $\lambda\in\Lambda$ (we adopt the usual physical convention that  probabilities vanish in regions of infinite potential).  For simplicity we assume $V(x)$ convex with superlogarithmic growth at infinity. It is known~\cite{Johansson98} that under these assumptions on $V(x)$ the Coulomb gas is supported in the thermodynamic limit on a bounded interval $[a,b]$. These assumptions on $V(x)$ are purely technical, and in fact the correct condition should be the stability of the one-cut configuration under small analytic perturbations of the external potential (this condition goes under the name of `off-criticality regime').

In the thermodynamic (large $N$) limit it is convenient to describe the positions $(\lambda_1,\dots,\lambda_N)$ of the gas particles (the eigenvalues of the $\beta$-ensemble) in terms of the empirical density of the Coulomb gas,
\be
\label{emp_den}
\rho_N(x)=\frac{1}{N}\sum_{i=1}^{N}\delta(x-\lambda_i)\ .
\ee
{The partition function of the gas may be written in terms of the empirical density in the following way. First, notice that for large $N$ the energy of the Coulomb gas admits an integral representation 
\be
-\frac{1}{2}\sum_{i\neq j}{\ln{|\lambda_i-\lambda_j|}}+N\sum_{i}{V(\lambda_i)}=N^2\EE[\rho_N]+o(N^2)
\ee
in terms of the mean-field energy functional $\EE$ defined as
\be
\EE[\sigma]=-\frac{1}{2}\iint_{x\neq y}\hspace{-2mm}\de\sigma(x)\de\sigma(y)\,\ln\abs{x-y}+\int\de\sigma(x)\,V(x)\ .\label{eq:functional}
\ee
Using this energy functional we have
\be
\mathcal{Z}_{N,\beta}=\int\mathcal{D}[\rho_N]\,\int\prod_{i=1}^N\de\lambda_i\, e^{-\beta N^2 \EE[\rho_N]+o(N^2)}\ ,
\ee
where the functional integration $\int\mathcal{D}[\rho_N]$ is restricted to the probability measures compatible with the Coulomb gas configuration $\{\lambda_i\}$.} A \emph{spectral linear statistics} (or \emph{linear statistics} for short) is defined  by
\begin{equation}
\label{specstat}
A_N=N^{-1}\sum_{i=1}^N f(\lambda_i)=\int\de x\,\rho_N(x)f(x),
\end{equation}
where $f(x)$ is a suitably chosen test function.  Sometimes~\eqref{specstat} is referred to as \emph{sum function} on the particles positions $\{\lambda_i\}$. The average of $A_N$ is given by
\be
\avg{A_N}=\int\de x\,\avg{\rho_N(x)}f(x).  \label{eq:avg_gen}
\ee
In a similar way one computes the variance of $A_N$, or in general the covariance between two linear statistics $A_N^{(1)}=N^{-1}\sum_{i=1}^N f_1(\lambda_i)$ and $A_N^{(2)}=N^{-1}\sum_{i=1}^N f_2(\lambda_i)$
as
\be
\Cov(A_N^{(1)},A_N^{(2)})=\int\de x\de y\,\Cov(\rho_N(x)\rho_N(y))f_1(x)f_2(y)\ . \label{eq:cov_gen}
\ee
The same reasoning can be extended to higher order cumulants. 
In general, the mixed cumulant $C_v(A_N^{(1)},A_N^{(2)},\dots,A_N^{(v)})$ of $v$ linear statistics (not necessarily distinct) can be recovered by integrating $A_N^{(j)}=N^{-1}\sum_{i=1}^N f_j(\lambda_i)$ against the cumulant  of the gas density at $v$ points $C_v(\rho_N(x_1),\rho_N(x_2),\dots,\rho_N(x_v))$, also known as \emph{$v$-point connected correlation function:}
\be
C_v(A_N^{(1)},A_N^{(2)},\dots,A_N^{(v)})=\int\left(\prod_{j=1}^{v}f_j(x_j)\de x_j\right) C_v(\rho_N(x_1),\rho_N(x_2),\dots,\rho_N(x_v)).
\ee
For $v=1$ and $v=2$ this corresponds to~\eqref{eq:avg_gen} and~\eqref{eq:cov_gen}, respectively.
In the asymptotic regime as $N\to \infty$, the connected correlation functions of the gas are generated by the free energy $-\log\mathcal{Z}_{N,\beta}$  by taking a suitable number of functional derivatives with respect to the external potential. More precisely, for every positive integer $v$:
\begin{equation}
C_v(\rho_N(x_1),\rho_N(x_2),\dots,\rho_N(x_v))=\left(-\frac{1}{\beta N^2}\right)^{v}\left[\prod_{i=1}^v\frac{\delta}{\delta V(x_i)}\right]\ln\mathcal{Z}_{N,\beta}\left[1+o(1)\right].\label{eq:lemma_der}
\end{equation}
For instance, the $1$-point and  $2$-point \emph{connected correlation functions} in the limit as $N\to \infty$ are given by
\begin{align}
\avg{\rho_N(x)}&=-\frac{1}{\beta N^2}\frac{\delta\ln\mathcal{Z}_{N,\beta}}{\delta V(x)}\left[1 + o(1)\right]\\
\Cov(\rho_N(x)\rho_N(y))&=\left(\frac{1}{\beta N^2}\right)^2\frac{\delta^2\ln\mathcal{Z}_{N,\beta}}{\delta V(x)\delta V(y)}\left[1 + o(1)\right]
\end{align}

Under the assumption that the limit  $\lim_{N \to \infty} \avg{\rho_N(x)} = \rho(x)$ exists,  from~\eqref{eq:lemma_der} we see that
\be
\lim_{N\to\infty}N^{2(v-1)}C_v(\rho_N(x_1),\rho_N(x_2),\dots,\rho_N(x_v))=\left(-\frac{1}{\beta}\right)^{v-1}\left[\prod_{i=1}^{v-1}\frac{\delta}{\delta V(x_i)}\right]\lim_{N\to\infty}\avg{\rho_N(x_v)} \label{eq:func_der_lim}
\ee 
exists too~\footnote{The last equality for $v=2$ was first recognized by Beenakker \cite{Beenakker93} in the random matrix approach to quantum transport  in chaotic cavities.}. Hence, in principle a complete answer to the general problem of computing spectral linear statistics in the large $N$ limit could be achieved if all functional derivatives with respect to the external confining potential were known. At present the complete solution to this problem is out of reach. Fortunately, in some circumstances we have a partial but explicit answer.  In the following we focus to first and second order effects:
\begin{align}
\rho(x)&=\lim_{N\to\infty}\avg{\rho_N(x)}\label{eq:dos}\\
\mathcal{K}(x,y)&=-\lim_{N\to\infty} N^2\Cov(\rho_N(x)\rho_N(y))=\frac{1}{\beta}\frac{\delta\rho(x)}{\delta V(y)}.\label{eq:2pk}
\end{align}

The quantity $\rho(x)$ is the \emph{density of states,} or \emph{equilibrium density,} while $\mathcal{K}(x,y)$ is known as \emph{smoothed two-point connected correlation kernel} (see \cite{BrezinZee93,Beenakker94} for details).  
Therefore, the limiting mean and covariances of smooth linear statistics (the lowest order cumulants) are 
\begin{align}
\lim_{N\to\infty}\avg{A_N}&=\int\de x\,\rho(x)f(x)\label{eq:mean}\\
\lim_{N\to\infty}N^2\Cov(A_N^{(1)},A_N^{(2)})&=-\int\de x\de y\,\mathcal{K}(x,y)f_1(x)f_2(x).\label{eq:cov}
\end{align}

From now on, we will restrict ourselves to one-cut ensembles, i.e. we will always assume that the potential $V(x)$ is convex and with superlogarithmic growth at infinity. Under these hypotheses, the density of states $\rho(x)$~\eqref{eq:dos} exists, is absolutely continuous and supported on a single bounded interval $\supp\rho=[a,b]$, ($a<b$), and is the unique probability measure that minimizes the { mean-field energy functional $\EE$ in~\eqref{eq:functional}.} We notice that $\beta$ does not appear in the minimization problem, and hence the density of states (and in particular its support) does not depend on $\beta$.
Remarkably, under the condition that the support is a single interval, $\rho(x)$ and $\mathcal{K}(x,y)$ have the closed expressions
\begin{align}
\rho(x)&=\frac{1}{\pi}\frac{\bm{1}_{a<x<b}}{\sqrt{(x-a)(b-x)}}
\left[1+\frac{1}{\pi}\,P\!\int_{a}^{b}\de y\frac{V'(y)\sqrt{(y-a)(b-y)}}{y-x}\right],\label{eq:one-point}\\
\mathcal{K}(x,y)&=\frac{1}{\beta\pi^2}\frac{\bm{1}_{a<x,y<b}}{\sqrt{(y-a)(b-y)}}
\frac{\de\,}{\de x} \frac{\sqrt{(x-a)(b-x)}}{x-y}\ ,
\label{eq:two-point}
\end{align}
where $P$ denotes Cauchy's principal value.

The first result~\eqref{eq:one-point} is usually referred to as Tricomi's formula~\cite{Tricomi}. The beautiful identity~\eqref{eq:two-point}
is a consequence of Tricomi's formula and~\eqref{eq:2pk}, and has been independently discovered by several authors ~\cite{Ambjorn90,BrezinZee93,Beenakker94,Kanzieper96,BrezinDeo99}. It is worth emphasizing that while the one-point density \eqref{eq:one-point} depends explicitly on the potential $V(x)$, the smoothed two-point connected correlation kernel \eqref{eq:two-point} enjoys a very surprising feature: it depends only on the edge points $a,b$ of the spectral density.  {The  dependence of $\mathcal{K}(x,y)$ on the details of the model only through the edges of the spectral density is often referred  to as to long-range (or macroscopic) universality. Conversely, the microscopic universality emerges on a scale $\Ord(1/N)$ of the interior points of the limiting support of $\rho(x)$ and is governed by the celebrated sine-kernel~\cite{BrezinZee93}. The universality of $\mathcal{K}(x,y)$ is a key observation that eventually leads to our main result, presented in next section.}

\section{Main Result}
\label{sec:main}
Let $X_{N,\kk}=N^{-1}\Tr\XX_N^{\kk}$ denote the moments of a generic one-cut $\beta$-ensemble of random matrices. For any fixed positive integers $\kk$ and $\ell$,  the limits
	\begin{align}
	\alpha_{\kk}&=\lim_{N\to\infty}\avg{X_{N,\kk}}\label{eq:def_avg}\\
  \frac{1}{\beta}\alpha_{\kk,\ell}&=\lim_{N\to\infty}N^2\Cov(X_{N,\kk},X_{N,\ell})\label{eq:def_cov}
	\end{align}
exist and  are given by
\begin{align}
\alpha_{\kk}&=\int_a^b\de x\,\rho(x)x^{\kk} \label{eq:mean_a} \\
\alpha_{\kk,\ell}&=\frac{1}{\pi^2}\,P\!\iint_a^b\de x\de y\,\sqrt{\frac{(x-a)(b-x)}{(y-a)(b-y)}}\frac{\kk}{y-x}y^{\ell}x^{\kk-1}, \label{eq:cov_struct}
\end{align}
respectively. The identities \eqref{eq:mean_a}-\eqref{eq:cov_struct} are a specialization of~\eqref{eq:mean}-\eqref{eq:cov}-\eqref{eq:two-point}. 
While the limiting average $\alpha_{\kk}$ depends explicitly on the probability measure that defines the ensemble through $V(x)$, and hence $\rho(x)$ (given in~\eqref{eq:one-point}), the covariance $(1/\beta)\alpha_{\kk,\ell}$ does not.   Indeed, the following theorem is \emph{universal,} as it depends only on the fact that the density of states is supported on a single bounded interval.
\begin{theorem}\label{thm:main}  
Let $\XX_N$ belong to a one-cut $\beta$-ensemble ($\beta>0$) and let $\supp \rho =[a,b]$ denote the support of the density of states.  Then, in the limit  $N\to \infty$ the generating function of the covariances~\eqref{eq:def_cov} is
\be
F_{[a,b]}(z,\zeta)=\frac{1}{\beta}\sum_{\kk,\ell=0}^{\infty}\alpha_{\kk,\ell}z^{\kk}\zeta^{\ell}=\frac{1}{\beta}\frac{z\zeta}{(z-\zeta)^2}\left[\frac{2abz\zeta-(a+b)(z+\zeta)+2}{2\sqrt{(1-az)(1-bz)(1-a\zeta)(1-b\zeta)}}-1\right]. \label{eq:main_res_gen}
\ee
\end{theorem}
This is the main result of this paper.

\begin{figure}[t]
\centering

\begin{tikzpicture}
[decoration={markings,
mark=at position 1cm with {\arrow[line width=1pt]{<}},
mark=at position 3.5cm with {\arrow[line width=1pt]{<}}
},scale=1.3
]
\draw[help lines,->, black, thick] (-.5,0) -- (.45,0)  (1.75,0) --(3,0)  coordinate (xaxis);
      \draw [dashdotted] (.45,0) -- (2,0);      
      \fill (1.75,0)  circle(0.05);
      \fill (0.45,0)  circle(0.05);
\draw [<->] (.05,-0.05) -- (.05,-.4);
     \draw [thick] (1.1,-0.05) -- (1.185,0.05);
     \draw [thick] (1.1,0.05) -- (1.185,-0.05);
\path[draw,line width=1.5pt,rounded corners, postaction=decorate] (0,0) +(.4,-0.4) --+(2,-0.4) --+(2,0.3) --+(.2,0.3)--+(.2,-0.4)--+(.4,-0.4);
\node[below] at (xaxis) {$\mathrm{Re} \zeta$};
\node at (.5,-.2) {$a$};
\node at (1.75,-.2) {$b$};
\node at (1.15,-.2) {$z$};
\node at (-.15,-.20) {$\epsilon'$};
\node at (1.15,.6) {\Large{$\Gamma_{\epsilon'}$}};
\node at (1,1.6) {\large{$\mathrm{i)}\,\,\displaystyle\lim_{\substack{\epsilon'\to0^+\\z\in(a,b)}}$}};
\end{tikzpicture} 
\hspace{1cm}
\begin{tikzpicture}
[decoration={markings,
mark=at position 1cm with {\arrow[line width=1pt]{<}},
mark=at position 3.5cm with {\arrow[line width=1pt]{<}}
},scale=1.3
]
\draw[help lines,->, black, thick] (-.5,.05) -- (.45,.05)  (1.75,.05) --(3,.05)  coordinate (xaxis);
      \draw [dashdotted] (.45,.05) -- (2,.05);      
      \fill (1.75,.05)  circle(0.05);
      \fill (0.45,.05)  circle(0.05);
\draw [<->] (.05,-0.03) -- (.05,-.4);
\path[draw,line width=1.5pt,rounded corners, postaction=decorate] (0,.0) +(.4,-0.4+.0) --+(2,-0.4+.0) --+(2,0.3+.0) --+(.2,0.3+.0)--+(.2,-0.4+.0)--+(.4,-0.4+.0);
\node[below] at (xaxis) {$\mathrm{Re} z$};
\node at (.5,-.2+.05) {$a$};
\node at (1.75,-.15) {$b$};
\node at (-.1,-.20) {$\epsilon$};
\node at (1.15,.6) {\Large{$\Gamma_{\epsilon}$}};
\node at (1,1.7) {\large{$\mathrm{ii)}\,\,\displaystyle\lim_{\substack{\epsilon\to0^+}}$}};
\end{tikzpicture} \label{fig:prescription}
\caption{Contours of integration and scheme of the prescription used to evaluate~\eqref{eq:integral2}. First the integral and the limit  $\epsilon'\to0^+$ in the $\zeta$-plane with $z$ inside the cut $[a,b]$. Then the integration in the $z$-plane.} 
\label{fig:contours}
\end{figure}
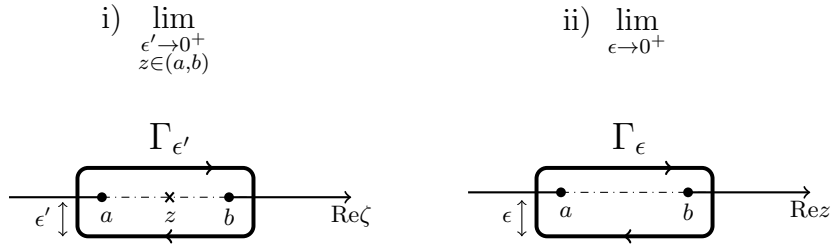

\begin{rmk} It is worthwhile to emphasize the following properties of $F_{[a,b]}(z,\zeta)$.
\begin{enumerate}
\item $F_{[a,b]}(z,\zeta)=F_{[a,b]}(\zeta,z)$, hence $\alpha_{\kk,\ell}=\alpha_{\ell,\kk}$, which expresses the fact that the covariance is a symmetric functional.
\item One can verify that $\lim_{z,\zeta\to0}F_{[a,b]}(z,\zeta)=0$. Therefore $\alpha_{\kk,\ell}=0$ whenever $\kk$ or $\ell$ are zero. This is a consequence of the fact that $X_{N,0}=1$ is a constant. Moreover, $F_{[a,b]}(z,\zeta)$ is continuous in $(z,\zeta)$ and analytic in each variable (keeping the other variable fixed) in a neighbourhood of the origin that does not intersect the cuts of the square roots in the denominator. If a function of complex variables is continuous and separately analytic in each variable, then it is also jointly analytic (Hartogs' theorem~\cite{Hartogs}); therefore,  $F_{[a,b]}(z,\zeta)$ is analytic in a neighbourhood of $(z,\zeta)=(0,0)$.
\item The limiting covariances are proportionals to $1/\beta$,  which is true whenever $V(x)$ is independent of $\beta$ (see, e.g.~{\rm \cite{Beenakker94,Forrester}}).
\item The generating function~\eqref{eq:main_res_gen} satisfies the \emph{homogeneity identity} $F_{[ta,tb]}(z,\zeta)=F_{[a,b]}(tz,t\zeta)$ for all real $t$. Therefore, under the map $(a,b)\mapsto(ta,tb)$ the limiting covariances transform as $\alpha_{\kk,\ell}\mapsto t^{\kk+\ell}\alpha_{\kk,\ell}$. This implies that the correlation coefficients $r_{\kk,\ell}=\alpha_{\kk,\ell}/\sqrt{\alpha_{\kk,\kk}\alpha_{\ell,\ell}}$ are invariant in absolute value. More precisely,
\be
r_{\kk,\ell}\mapsto(\mathrm{sign}(t))^{\kk+\ell}r_{\kk,\ell}\quad \text{as} \quad (a,b) \mapsto (ta,tb).
\ee
\item When the support of the equilibrium density is centred at the origin, i.e $\supp \rho =[-b,b]$ with $b>0$, then the generating function simplifies considerably:
\be
\label{symmetricF}
F_{\pm b}(z,\zeta) =\frac{1}{\beta}\frac{z\zeta}{(z-\zeta)^2}\left[\frac{1 - b^2 z\zeta}{\sqrt{(1-b^2z^2)(1-b^2\zeta^2)}}-1\right]. 
\ee
Furthermore, it satisfies $F_{\pm b}(z,\zeta)=F_{\pm b}(-z,-\zeta)$; as a consequence $\alpha_{\kk,\ell}=0$ whenever $\kk\neq\ell\mod2$.
\item  The support $[a,b]$ of the density of states can always be centred at the origin by a shift $\lambda \mapsto\lambda - m $, where $m=\left(a + b\right)/2$. This corresponds to a translation in the matrix model $\XX_N \mapsto\XX_N - m I_N $ where $I_N$ is the identity matrix. It is known that the 2D Coulomb gas interaction is invariant under conformal transformations~{\rm\cite{CundenVivo}}. In other words, the covariance structure depends uniquely on the fact that the support is a single bounded interval. The translation $\XX_N \mapsto \XX_N - m I_N$ induces the change of variables $ x \mapsto x - m$ and  $y \mapsto y-m$ into the integral in the  r.h.s. of~\eqref{eq:cov_struct}, which leads to the relation
\be
\label{eq:cov_struct2}
\alpha_{\kk,\ell}=m^{\kk+\ell}\sum_{p=0}^{\kk}\sum_{q=0}^{\ell}\binom{\kk}{p}\binom{\ell}{q}
\left(\frac{L}{2m}\right)^{p+q}\alpha^{\G}_{p,q}
\ee
where $m=(a+b)/2$, $L=(b-a)/2$ and $\alpha^{\G}_{p,q}$ are the covariances of an equilibrium density supported on $[-2,2]$ (e.g. of the Gaussian ensemble). The $\alpha^{\G}_{p,q}$ are explicitly given in~\eqref{eq:covG}.  
Thus, \eqref{eq:cov_struct2} gives a tool to evaluate the limiting covariances for an arbitrary one-cut $\beta$-ensemble.  This is somehow analogous to the usual process of centring the variables in the theory of probability.
\item If $\alpha_{\kk,\ell}\in\N$ for all $\kk,\ell$ then $\max(|a|,|b|)>1$. This is a necessary condition to have positive integers $\alpha_{\kk,\ell}$.
\end{enumerate}\label{rmk1}
\end{rmk}
In what follows we present the proof of~\eqref{eq:main_res_gen}.
{As  in~\cite{Cunden14}, the real integral~\eqref{eq:cov_struct} can be lifted} to a double integral in the complex plane,
 \be
\alpha_{\kk,\ell}=\lim_{\epsilon\to0^+}\lim_{\epsilon'\to0^+}\frac{1}{\pi^2}\,\int_{\Gamma_{\epsilon}}\!\!\de z\int_{\Gamma_{\epsilon'}}\!\!\de\zeta\,\sqrt{\frac{(z-a)(z-b)}{(\zeta-a)(\zeta-b)}}\frac{\kk}{\zeta-z}\zeta^{\ell}z^{\kk-1},\label{eq:integral2}
\ee
where $\Gamma_{\epsilon}$, $\Gamma_{\epsilon'}$ are two clockwise oriented contours enclosing the cut $[a,b]$ at distance $\epsilon$ and $\epsilon'$ respectively (see Fig.~\ref{fig:contours}). 
{The first integration with respect to $\zeta$ is performed with $z$ fixed inside the cut. In this way the singularity $\zeta=z$ becomes irrelevant. One verifies that the result of the first integration (a function of $z$) is analytic in a neighbourhood of the cut and hence integrable on the contour $\Gamma_{\epsilon}$ enclosing the cut. With this prescription in mind, for $\kk,\ell\geq0$, we get using residues}
\begin{align}
\alpha_{\kk,\ell}&=\kk\times\Res\left(\sqrt{\frac{(z-a)(z-b)}{(\zeta-a)(\zeta-b)}}\frac{1}{z-\zeta}\zeta^{\ell}z^{\kk-1};\zeta=\infty,z=\infty\right) \label{eq:cov_structure}\\
&=\kk\times\Res\left(\sqrt{\frac{(1-az)(1-bz)}{(1-a\zeta)(1-b\zeta)}}\frac{\zeta}{z-\zeta}\frac{1}{\zeta^{\ell+1}}\frac{1}{z^{\kk+1}};\zeta=0,z=0\right)\label{eq:cov_last_step}.
\end{align}
From~\eqref{eq:cov_last_step} we have
\begin{align}
\alpha_{\kk,\ell}&=\kk\times\text{coefficient of $z^{\kk}\zeta^{\ell}$ in the expansion of} \left(\sqrt{\frac{(1-az)(1-bz)}{(1-a\zeta)(1-b\zeta)}}\frac{\zeta}{z-\zeta}\right)\text{ about $(z,\zeta)=(0,0)$}\label{eq:cov_last_step2}\\
&=\text{coefficient of $z^{\kk}\zeta^{\ell}$ in the expansion of } z\,\partial_z\left(\sqrt{\frac{(1-az)(1-bz)}{(1-a\zeta)(1-b\zeta)}}\frac{\zeta}{z-\zeta}\right)\text{ about $(z,\zeta)=(0,0)$}.\label{eq:cov_last_step3}
\end{align}
In the last line we have used the observation that if $C(z)=\sum_{\kk}c_{\kk}z^{\kk}$, then the generating function of  $\kk c_{\kk}$ is $z\partial_z C(z)$.
Hence, the generating function $F(z,\zeta)=(1/\beta)\sum_{\kk,\ell\geq0}{\alpha_{\kk,\ell}z^{\kk}\zeta^{\ell}}$ is given by
\be
F_{[a,b]}(z,\zeta)=\frac{1}{\beta}\times\text{analytic part at $(z,\zeta)=(0,0)$ of the function }z\,\partial_z\left(\sqrt{\frac{(1-az)(1-bz)}{(1-a\zeta)(1-b\zeta)}}\frac{\zeta}{z-\zeta}\right).
\ee
Carrying out the differentiation and isolating the analytic part, one gets the final expression~\eqref{eq:main_res_gen}.

We also have an alternative derivation of~\eqref{eq:main_res_gen}, which relies on the chain of loop equations for the $v$-point resolvent 
\be
W(z_1,\dots, z_v)=C_v\left(\frac{1}{N}\tr\frac{1}{z_1-\XX_N},\dots,\frac{1}{N}\tr\frac{1}{z_v-\XX_N}\right).
\ee
When the equilibrium density has square root singularities at the edges of the support (soft edges) these quantities admit a power expansion in $1/N$ (we refer to~\cite{Chekhov06,Guionnet13} for a more detailed exposition) and, under certain hypotheses,  one can compute the $v$-point resolvents recursively order by order by solving a hierarchy of  loop equations (for an algorithmic approach see~\cite{Borot11}). Within this formalism, the generating function~\eqref{eq:main_res_gen} can be expressed as $F_{[a,b]}(z,\zeta)=z\zeta W^{(0)}_2(1/z,1/\zeta)$ where $W^{(0)}_2$ is the leading order in $1/N$ of the $2$-point resolvent. Using the notation in~\cite{Borot11}, we have
\begin{align}
F_{[a,b]}(z,\zeta)& =\frac{2}{\beta z\zeta}\,\omega\left(j_{a,b}^{-1}(1/z),j_{a,b}^{-1}(1/\zeta)\right),\label{eq:Juk1}\\
\omega(x,y)& =\frac{1}{\left(j_{a,b}'(x)-j_{a,b}'(y)\right)^2}\frac{1}{\left(xy-1\right)^2},\label{eq:Juk2}
\end{align}
{where $j_{a,b}$ is the \emph{Joukowski transformation} defined as $j_{a,b}(z)=\frac{a+b}{2}+\frac{a-b}{4}\left(z+\frac{1}{z}\right)$ (in~\eqref{eq:Juk1}-\eqref{eq:Juk2} $j_{a,b}^{-1}$ and $j_{a,b}'$ denote the inverse map and the derivative of $j_{a,b}$, respectively).}  

\section{Applications to the classical $\beta$-ensembles}\label{sec:applications}
In this Section we focus on the classical $\beta$-ensembles: the Gaussian, Wishart and Jacobi $\beta$-ensembles. 

Fluctuations of traces in the Gaussian and Wishart $\beta$-ensembles were considered by Dumitriu and Edelman~\cite{Dumitriu06} in the framework of the tridiagonal realization of $\beta$-ensemble that they introduced~\cite{Dumitriu02} (for $\beta=1,2$ the tridiagonalization of random matrices had already appeared earlier in~\cite{Trotter84} and \cite{Silverstein85}). Using a combinatorial technique simplified by the sparseness of their matrix models, they managed to obtain formulae for the averages and covariances of the moments (among other results). Their formulae are insensitive to the particular value of $\beta>0$ (since $\beta$ appears as just a parameter in the distribution of off-diagonal terms). Later, Dumitriu and Paquette~\cite{Dumitriu12} investigated the very same statistics on the Jacobi ensemble. The matrix realization of this ensemble  is not tridiagonal~\cite{KN04}; this turns out to be a serious obstruction to obtain explicit formulae.

Here we show that our approach allows to treat these cases on the same footing: \emph{neither} the specific matrix model \emph{nor} the concrete matrix representation do matter. We will recover previous results, and produce new self-contained formulae for the limiting covariances of the Wishart and Jacobi $\beta$-ensemble as particular cases of the general formula \eqref{eq:main_res_gen}.

\subsection{Gaussian ensemble and symmetric intervals $[a,b]=[-L,L]$}
\label{sub:Gaussian}
The density of states of the $\beta$-Gaussian ensemble is symmetric with respect to the origin.  It is, therefore, a natural place to start. The eigenvalues of $\beta$-Gaussian matrices  $\G_N$ are distributed according to 
\be
\de\mathbb{P}_{N,\beta}(\{\lambda_k\})\propto |\Delta(\{\lambda_k\})|^{\beta}\prod_{k=1}^Ne^{-N\beta\lambda_k^2/4}\de\lambda_k, 
\label{jpdfbetagauss}
\ee
where $\Delta(\{\lambda_k\})=\prod_{j<k}(\lambda_k-\lambda_j)$ is the Vandermonde determinant. Eq.~\eqref{jpdfbetagauss} is easily recognized as the canonical measure of a 2D Coulomb gas in a quadratic potential $V(x)=x^2/4$.
When $\beta=1,2,4$ \eqref{jpdfbetagauss} is the joint probability density function of the eigenvalues of real symmetric (\textsf{GOE}), complex Hermitian (\textsf{GUE}) and quaternion self-dual (\textsf{GSE}) gaussian random matrices, respectively.

For any $\beta>0$ the average of the empirical density~\eqref{emp_den} of the {eigenvalues} $\left \{\lambda_i\right \}$ converges in expectation to  Wigner's semicircle law,
\be
\rho_{\G}(x)=\frac{1}{2\pi}\sqrt{4-x^2}\,{\bm 1}_{|x|<2}.
\ee
Therefore, the averages of the  moments $\alpha_{\kk}^{\G}=\lim_{N\to\infty}\avg{G_{N,\kk}}$ converge to the Catalan numbers
\be
\alpha_{\kk}^{\G}=\int\de x\rho_{\G}(x) x^{\kk}=
\begin{cases}
\displaystyle\frac{2}{\kk+2}\binom{\kk}{\frac{\kk}{2}}\quad &\text{for}\,\, \kk=0\hspace{1mm}\text{even}\\
0\quad &\text{for}\,\, \kk=0\hspace{1mm}\text{odd}. 
\end{cases}\label{eq:alphak}
\ee 
The limiting density $\rho_{\G}$ is supported on the interval $[a,b]=[-2,2]$. Thus, formula \eqref{eq:main_res_gen} reads
\be
F_{\pm2}(z,\zeta)=\frac{1}{\beta}\frac{z\zeta}{(z-\zeta)^2}\left[\frac{1-4z\zeta}{\sqrt{(1-4z^2)(1-4\zeta^2)}}-1\right]\ . \label{eq:main_res_G}
\ee

The coefficients $\alpha_{\kk,\ell}^{\G}$ of this double series can be first extracted by taking $k$th and $\ell$th order derivatives of \eqref{eq:main_res_G} symbolically, and recognizing an easy pattern. This leads eventually to the well-supported ansatz

    \be
\alpha_{\kk,\ell}^{\G}=\lim_{N\to\infty}\beta N^2\Cov\left(G_{N,\kk},G_{N,\ell}\right)=
\begin{cases}
\displaystyle\frac{4\kk\ell}{\kk+\ell}\binom{\kk-1}{\floor*{\frac{\kk}{2}}}\binom{\ell-1}{\floor*{\frac{\ell}{2}}}\quad &\text{if}\,\, \kk=\ell\hspace{-1mm}\mod2\\
0\quad &\text{if}\,\, \kk\neq\ell\hspace{-1mm}\mod2.
\end{cases}\label{eq:covG}
\ee
where $\floor*{x}$ is the largest integer smaller than or equal to $x$. Then, it remains to \emph{prove} that the guessed form \eqref{eq:covG} is indeed correct by resumming the series $\sum_{\kappa,\ell}\alpha_{\kappa,\ell}^{\mathcal{G}}z^\kappa \zeta^\ell$ - this is outlined in Appendix \ref{appB}.

The first values of these covariances are given explicitly in Table \ref{tab:num}. We mention that~\eqref{eq:covG} can be tracked back to a work by Tutte~\cite{Tutte62} in combinatorics (see Appendix~\ref{app:review}). 

\begin{table}[h]
\[\lim_{N\to\infty}N^2\Cov\left(G_{N,\kk},G_{N,\ell}\right)=\frac{1}{\beta}
\left[
\begin{array}{cccccccc}
 2 & 0 & 6 & 0 & 20 & 0 & 70 & 0 \\
 0 & 4 & 0 & 16 & 0 & 60 & 0 & 224 \\
 6 & 0 & 24 & 0 & 90 & 0 & 336 & 0 \\
 0 & 16 & 0 & 72 & 0 & 288 & 0 & 1120 \\
 20 & 0 & 90 & 0 & 360 & 0 & 1400 & 0 \\
 0 & 60 & 0 & 288 & 0 & 1200 & 0 & 4800 \\
 70 & 0 & 336 & 0 & 1400 & 0 & 5600 & 0 \\
 0 & 224 & 0 & 1120 & 0 & 4800 & 0 & 19600
\end{array}
\right]
\]
\caption{A block of limiting covariance structure ($1\leq\kk,\ell\leq8$) for the $\beta$-Gaussian ensemble from \eqref{eq:covG}.}
\label{tab:num}
\end{table}

More generally, using the homogeneity property of $F_{[a,b]}(z,\zeta)$ (see Remark~\ref{rmk1}), for \emph{any} one-cut $\beta$ ensemble $\XX_N$ whose limiting spectral density is supported on a single symmetric interval $[-L,L]$ we have
    \be
\lim_{N\to\infty}N^2\Cov\left( X_{N,\kk},X_{N,\ell}\right)=\frac{1}{\beta}
\begin{cases}
\displaystyle\left(\frac{L}{2}\right)^{\kk+\ell}\frac{4\kk\ell}{\kk+\ell}\binom{\kk-1}{\floor*{\frac{\kk}{2}}}\binom{\ell-1}{\floor*{\frac{\ell}{2}}}\quad &\text{if}\,\, \kk=\ell\hspace{-1mm}\mod2\\
0\quad &\text{if}\,\, \kk\neq\ell\hspace{-1mm}\mod2
\end{cases}\label{eq:cov[-L,L]}\qquad\text{if } [a,b]=[-L,L].
\ee
For $L=1$ this expression agrees with the formula derived by Dumitriu and Edelman~\cite{Dumitriu06} with a much more complicated combinatorial proof.~\footnote{In~\cite{Dumitriu06} a different choice of the scaling of the eigenvalues leads to an equilibrium density supported on $[-1,1]$.} 

It is worth emphasizing that all the ensembles with a symmetric support of the equilibrium density share the same $L$- and $\beta$-independent correlation matrix (see Remark~\ref{rmk1}),
\be
r_{\kk,\ell}=\frac{\alpha_{\kk,\ell}}{\sqrt{\alpha_{\kk,\kk}\alpha_{\ell,\ell}}}=
\begin{cases}
\displaystyle2\frac{\sqrt{\kk\ell}}{\kk+\ell}\quad &\text{if}\,\, \kk=\ell\hspace{-1mm}\mod2\\
0\quad &\text{if}\,\, \kk\neq\ell\hspace{-1mm}\mod2
\end{cases}\label{eq:r[-L,L]}\qquad\text{if } [a,b]=[-L,L].
\ee

\subsection{Wishart ensemble and intervals $[a,b]=[0,2L]$}

The same direct method that we used for the Gaussian $\beta$-ensemble can be applied to the $\beta$-Wishart (Laguerre) ensemble of positive semi-definite random matrices $\W_N\geq0$, characterized by a joint probability density of the eigenvalues 
\be
\de\mathbb{P}_{N,\beta}(\{\lambda_k\})\propto |\Delta(\{\lambda_k\})|^{\beta}\prod_{k=1}^N\lambda_{k}^{\beta(c-1)N}e^{-\frac{\beta }{2}N\lambda_k}\theta(\lambda_k)\,\de\lambda_k\ ,\qquad(c\geq1).
\label{jpdfwisharteig}
\ee
For $\beta=1,2,4$, this ensemble can be realized as $\W_N=\frac{1}{N}\G_{N\times M}\G^\dagger_{N \times M}$, where $\G_{N\times M}$ denotes a Gaussian $N\times M$ matrix with {i.i.d.} real, complex or quaternion entries, respectively. The parameter $c\geq1$ in \eqref{jpdfwisharteig} is the ratio $M/N$, which is assumed to remain constant as $M,N\to\infty$. We notice that~\eqref{jpdfwisharteig} is the canonical measure of a 2D Coulomb gas in the (non-polynomial) external potential $V(x)=x/2-(c-1)\ln x$ for nonnegative $x$ and infinite on the negative half-line.

The average of the empirical spectral density of this ensemble converges to the Mar\v{c}enko-Pastur law with parameter $c\geq1$:
\be
\label{eq:marcenko_pasteur}
\rho_{\W}(x)=\frac{1}{2\pi x}\sqrt{(x-a)(b-x)}\,{\bm 1}_{x\in(a,b)}, \qquad\text{with }\quad [a,b]=[(1-\sqrt{c})^2,(1+\sqrt{c})^2]\ .
\ee
The limit on average $\alpha^{\W}_{\kk}=\lim_{N\to\infty}\avg{W_{N,\kk}}$ of the moments $W_{N,\kk}=N^{-1}\tr\W_N^{\kk}$ exists and is described in terms of the  \emph{Narayana numbers:}
\be
\alpha^{\W}_{\kk}=\sum_{p=1}^{\kk}c^{p-1}\mathrm{Nar}(\kk,p), \qquad \mathrm{Nar}(\kk,p)=\frac{1}{p}\binom{\kk}{p-1}\binom{\kk-1}{p-1}.
\ee
Formula~\eqref{eq:cov_struct2} combined with~\eqref{eq:covG} gives 
\be
\alpha^{\W}_{\kk,\ell}=\lim_{N\to\infty}\beta N^2\Cov\left(W_{N,\kk},W_{N,\ell}\right)=4(1+c)^{\kk+\ell}\!\!\sum_{\substack{0\leq p\leq\kk\\0\leq q\leq \ell\\p=q\bmod2}}\!\!\left(\frac{\sqrt{c}}{1+c}\right)^{p+q}\frac{pq}{p+q}\binom{\kk}{p}\binom{\ell}{q}\binom{p-1}{\floor*{\frac{p}{2}}}\binom{q-1}{\floor*{\frac{q}{2}}}.
\label{eq:neat}
\ee
Few values of $\alpha^{\W}_{\kk}$ are reported in Table~\ref{tab:numW}.

The $\beta$-Wishart ensemble was also investigated by Dumitru and Edelman~\cite{Dumitriu06} within the framework of the tridiagonal realization of the ensemble. Their proof requires an ad hoc  combinatorial analysis, which is different from the $\beta$-Gaussian ensemble. Their formula (see~\cite{Dumitriu06}, Claim~3.15.2) is considerably more involved than~\eqref{eq:neat}, and besides, we found that it contains few misprints.

When the support is the interval $[0,2L]$, the generating function of the $\alpha_{\kk,\ell}$ simplifies drastically
\be
F_{[0,2L]}(z,\zeta)=\frac{1}{\beta}\frac{z\zeta}{(z-\zeta)^2}\left[\frac{1-L(z+\zeta)}{\sqrt{(1-2Lz)(1-2L\zeta)}}-1\right]. \label{eq:main_res_Wsym}
\ee
Extracting the Taylor coefficients of this formula involves lengthy calculations. 
{However, for $c=1$ the moments of Wishart matrices are related in a simple way to the moments of Gaussian matrices and therefore one obtains a self-contained formula for the covariances,}
    \be
\lim_{N\to\infty}N^2\Cov\left( X_{N,\kk},X_{N,\ell}\right)=\frac{1}{\beta}
\displaystyle\left(\frac{L}{2}\right)^{\kk+\ell}\frac{4\kk\ell}{\kk+\ell}\binom{2\kk-1}{\kk}\binom{2\ell-1}{\ell},\qquad\text{if } [a,b]=[0,2L].\label{eq:cov[0,L]}
\ee
Setting $2L=4$ we get the covariance structure $\alpha_{\kappa,\ell}^{\W}$ of the symmetric ($c=1)$ Wishart ensemble.
\begin{table}[t]
\[
\lim_{N\to\infty}N^2\Cov\left(W_{N,\kk},W_{N,\ell}\right)=
\frac{1}{\beta}\left[
\begin{array}{ccc}
 2 c & 4 \left(c+c^2\right) & 6 \left(c+3 c^2+c^3\right) \\
 4 \left(c+c^2\right) & 4 \left(2 c+5 c^2+2 c^3\right) & 12 \left(c+5 c^2+5 c^3+c^4\right) \\
 6 \left(c+3 c^2+c^3\right) & 12 \left(c+5 c^2+5 c^3+c^4\right) & 6 \left(3 c+24 c^2+46 c^3+24 c^4+3 c^5\right)
\end{array}
\right]
\]
\caption{Block of the limiting covariance structure ($1\leq \kk,\ell\leq3$) of the $\beta$-Wishart ensemble from~\eqref{eq:neat}. 
}
\label{tab:numW}
\end{table}

\begin{figure}[t]
\centering
\includegraphics[width=.75\columnwidth]{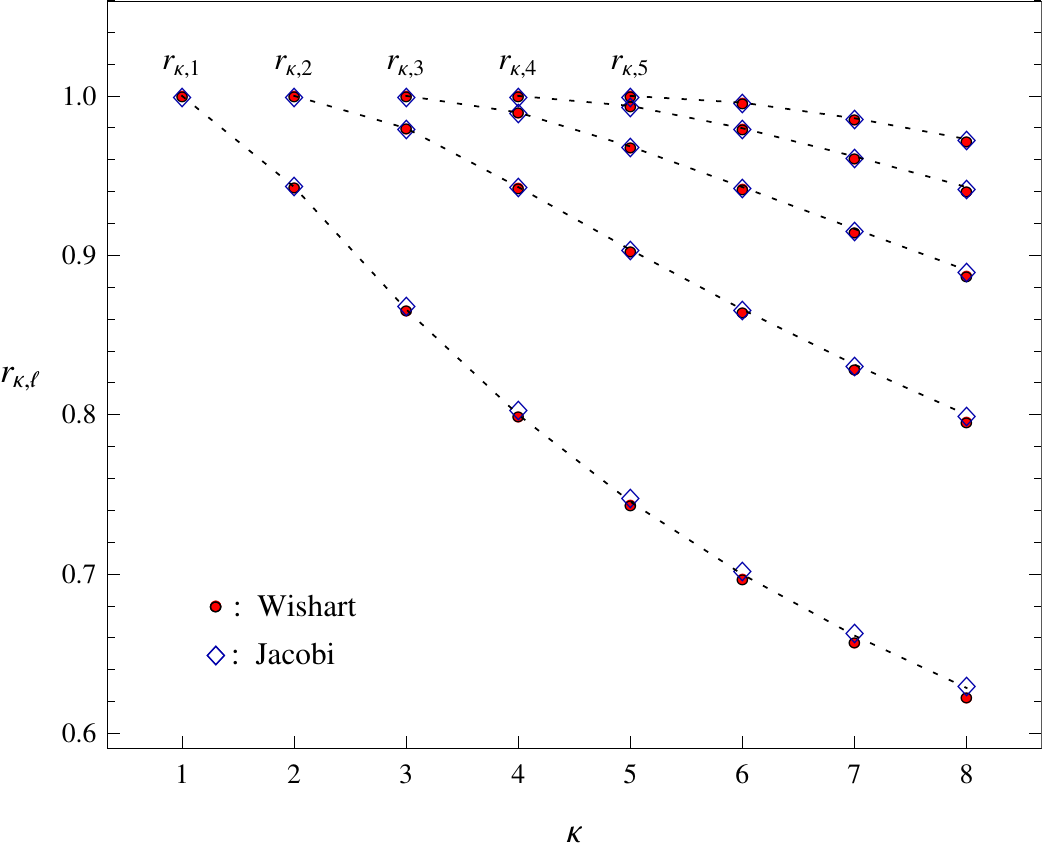}
\caption{Limiting correlations $r_{\kk,\ell}=\alpha_{\kk,\ell}/\sqrt{\alpha_{\kk,\kk}\alpha_{\ell,\ell}}$ of the moments for the Wishart ensemble $\W_N$ with $c=1$ and the Jacobi ensemble $\J_N$ with $\gamma_1=\gamma_2=0$. The densities of states $\rho_{\W}$ and $\rho_{\J}$ are supported on $[a,b]=[0,4]$ and $[a,b]=[0,1]$, respectively. According to Remark~\ref{rmk1}, the correlation coefficients $r_{\kk,\ell}$ are the same for these ensembles. Here we show a comparison of {the} prediction~\eqref{eq:r[0,L]} with numerical simulations. The size of the matrices $\W_N$ and $\J_N$ is $N=50$, the sample size is $n=10^{4}$, $\kk=1,...,8$ and $\ell=1,...,5$. Dotted lines connecting the theoretical values are depicted to guide the eye.}
\label{fig:numerics}
\end{figure}

The correlation coefficients are independent of the length of the interval and are given 
\be
r_{\kk,\ell}=\frac{\alpha_{\kk,\ell}}{\sqrt{\alpha_{\kk,\kk}\alpha_{\ell,\ell}}}=2\frac{\sqrt{\kk\ell}}{\kk+\ell},\qquad\text{if } [a,b]=[0,2L].\label{eq:r[0,L]}
\ee
In Fig.~\ref{fig:numerics} we report Monte Carlo simulations of correlation coefficients for ensembles of type $[0,2L]$. From~\eqref{eq:cov[0,L]} we can derive the large $\kk,\ell$ behaviour of the covariances
\be
\lim_{N\to\infty}N^2\Cov\left( X_{N,\kk},X_{N,\ell}\right)\sim \frac{(2L)^{\kk+\ell}}{\beta \pi}\frac{\sqrt{\kk\ell}}{\kk+\ell}\qquad\text{if $[a,b]=[0,2L]$,}\label{asymp_covar} 
\ee
in agreement with an earlier prediction of two of the authors of this paper~\cite{CundenVivo}.

\subsection{Jacobi ensemble}
As last example we consider the $\beta$-Jacobi ensemble. These matrices satisfy the constraint $0\leq\J_N\leq I$ and their eigenvalues $0\leq\lambda_k\leq1$ have joint probability density function
\be
\de\mathbb{P}_{N,\beta}(\{\lambda_k\})\propto   |\Delta(\{\lambda_k\})|^{\beta}\prod_{k=1}^N\lambda_{k}^{\frac{\beta}{2}N\gamma_1}(1-\lambda_{k})^{\frac{\beta}{2}N\gamma_2}\theta(\lambda_k(1-\lambda_k))\,\de\lambda_k\ ,\qquad(\gamma_{1,2}\geq0).
\ee

The density of states of this ensemble is 
\be
\rho_{\J}(x)=\frac{\gamma_1+\gamma_2+2}{2\pi x(1-x)}\sqrt{(x-a)(b-x)}\,{\bm 1}_{x\in(a,b)}.
\ee
It is compactly supported on the interval $[a(\gamma_1,\gamma_2),b(\gamma_1,\gamma_2)]$, whose edges are
\be
a(\gamma_1,\gamma_2) = \left[\frac{\sqrt{\gamma_2+1}- \sqrt{(\gamma_1+1)(\gamma_1+\gamma_2+1)}}{\gamma_1+\gamma_2+2}\right]^2, \qquad b(\gamma_1,\gamma_2)= \left[\frac{\sqrt{\gamma_2+1}+ \sqrt{(\gamma_1+1)(\gamma_1+\gamma_2+1)}}{\gamma_1+\gamma_2+2}\right]^2.\label{eq:edgesJ}
\ee
By inserting~\eqref{eq:edgesJ} into~\eqref{eq:main_res_gen} one obtains the generating function of the limiting covariances  of $J_{N,\kk}=\tr\J_N^{\kk}$ for any value of $\gamma_{1,2}\geq 0$. The $\alpha_{\kk,\ell}^{\J}$'s can then be evaluated explicitly via~\eqref{eq:cov_struct2} as 
\be
\alpha^{\J}_{\kk,\ell}=\left[\frac{\gamma_1^2+\gamma_1\gamma_2+2(\gamma_1+\gamma_2+1)}{(\gamma_1+\gamma_2+2)^2}\right]^{\kk+\ell}\sum_{\substack{0\leq p\leq\kk\\0\leq q\leq \ell}}\left[\frac{\sqrt{(\gamma_1+1)(\gamma_2+1)(\gamma_1+\gamma_2+1)}}{\gamma_1^2+\gamma_1\gamma_2+2(\gamma_1+\gamma_2+1)}\right]^{p+q}\binom{\kk}{p}\binom{\ell}{q}
\alpha^{\G}_{p,q},
\label{eq:covJ}
\ee
where $\alpha^{\G}_{p,q}$ is given in~\eqref{eq:covG}.
Furthermore, in the particular case $\gamma_{1}=\gamma_{2}=0$, we have $[a,b]=[0,1]$; hence, using formula~\eqref{eq:cov[0,L]} we arrive at
 \be
\alpha_{\kk,\ell}^{\J}=\lim_{N\to\infty}\beta N^2\Cov\left( J_{N,\kk},J_{N,\ell}\right)=
\displaystyle4^{1-\kk-\ell}\frac{\kk\ell}{\kk+\ell}\binom{2\kk-1}{\kk}\binom{2\ell-1}{\ell}\qquad\text{if } \gamma_{1,2}=0 \quad ([a,b]=[0,1]).\label{eq:Jsymm}
\ee
Few values are reported in Table~\ref{tab:numJ}.

Dumitriu and Paquette~(see \cite{Dumitriu12}, Eq. (23)) derived an integral formula for $\alpha_{\kk,\ell}^{\J}$ that can only be evaluated numerically; one can then check that the numerical integration gives the same values as formula~\eqref{eq:covJ}. 
Their proof, however, does not extend to the limiting values $\gamma_{1,2}=0$, which are covered by ours.
\begin{table}[h]
\[
\lim_{N\to\infty}N^2\Cov\left( J_{N,\kk},J_{N,\ell}\right)=\frac{1}{\beta}
\left[
\begin{array}{ccccc}
 \displaystyle\frac{1}{8} &\displaystyle \frac{1}{8} & \displaystyle\frac{15}{128} &\displaystyle \frac{7}{64} &\displaystyle \frac{105}{1024} \\\\
 \displaystyle \frac{1}{8} & \displaystyle\frac{9}{64} & \displaystyle\frac{9}{64} & \displaystyle\frac{35}{256} & \displaystyle\frac{135}{1024} \\\\
 \displaystyle \frac{15}{128} & \displaystyle\frac{9}{64} & \displaystyle\frac{75}{512} &\displaystyle \frac{75}{512} & \displaystyle\frac{4725}{32768} \\\\
 \displaystyle \frac{7}{64} &\displaystyle \frac{35}{256} & \displaystyle\frac{75}{512} & \displaystyle\frac{1225}{8192} & \displaystyle\frac{1225}{8192} \\\\
  \displaystyle\frac{105}{1024} & \displaystyle\frac{135}{1024} &\displaystyle \frac{4725}{32768} &\displaystyle \frac{1225}{8192} & \displaystyle\frac{19845}{131072}
\end{array}
\right]
\]
\caption{A block of the limiting covariance structure ($1\leq \kk,\ell\leq5$) of the Jacobi ensemble for $\gamma_{1,2}=0$ from~\eqref{eq:Jsymm}.}
\label{tab:numJ}
\end{table}

\section{Conclusions}
\label{sec:concl}
Using a Coulomb gas approach, we have derived a formula~\eqref{eq:main_res_gen} for the generating function of the covariances of the  moments in the thermodynamic limit $N\to \infty$ for one-cut $\beta$-ensembles of random matrices.   This result is universal, as it depends only on the support of the limiting spectral density, but not on the potential $V(x)$ that defines the ensemble, nor on the inverse temperature $\beta$; this is a direct consequence of the universality of the smoothed two-point spectral correlation function.  Given a $\beta$-ensemble whose equilibrium density has support $[a,b]$, the covariances of the moments can always be expressed in terms of a set of \emph{centred} covariances, i.e. corresponding to a limiting density with support centred at the origin. This allows us to write an explicit self-contained formula~\eqref{eq:cov_struct2} for the limiting covariances of \emph{any} one-cut $\beta$-ensemble.  Since the proof is based on the Dyson Coulomb gas analogy it is independent of the particular matrix realization of the ensemble and as such is simpler and more transparent than methods previously available.

The classical $\beta$-Gaussian, $\beta$-Wishart and $\beta$-Jacobi ensembles can be treated in a unified way. In this way we recover results previously established~\cite{Dumitriu02,Dumitriu06} with proofs based on sparse matrix realization of $\beta$-ensembles.   In addition we have proved new formulae for the $\beta$-Wishart and the $\beta$-Jacobi ensembles.
{We mention that the issues of second order cumulants and joint behavior of spectral linear statistics have recently attracted attention in the context of the theory of quantum transport in mesoscopic cavities~\cite{Grabsch14,Cunden14,Cunden15} as well as in practical spectral inference~\cite{Drogosz15,CundenVivo2}.}

We conclude by mentioning several related open problems. The challenge posed by higher cumulants is still open and the promising features of the 2D Coulomb gas approach suggest that the same line of reasoning could be pursued to investigate the universality of the $v$-point correlation kernel for $v>2$. Concerning the first and second order cumulants, it is natural to look for the next to leading term corrections. In~\cite{Johansson98}, Johansson showed how to compute the subleading terms for the averages at any $\beta>0$ (for $V$ polynomial). We are not aware of similar results for the $o(N^{-2})$ corrections of the covariance structures. {Coming back to leading order effects, it would be interesting to obtain an analogue of~\eqref{eq:main_res_gen} for multi-cut $\beta$-ensembles where the universality of the two-point smoothed kernel is more delicate~\cite{Albeverio01,Akemann96,Bonnet00}.}  As a last remark, we point out that the Coulomb gas approach is tailored to invariant ensembles. Wigner matrices (other than Gaussian) escape from the unitary invariance and this is a serious problem when discussing the $v$-point correlation kernels. For Wigner matrices it is known that, while the density of states is universal (the Wigner law), the higher-order {correlation kernels} are somewhat less universal~\cite{Zee96} (they depend on more details of the entries distribution). The solution to these problems may well uncover yet another layer of ``universality" features of ``classical" \textsf{RMT}, a field from which many lessons are surely still to be learnt.

\section*{Acknowledgments}
FDC and FM are grateful to G. Borot for enlightening discussions on the loop equations technique and for suggestion of a few relevant references. This work was partially supported by EPSRC Grant number EP/L010305/1. FDC acknowledges partial support of Gruppo Nazionale di Fisica Matematica GNFM-INdAM. No empirical or experimental data were created during this study.
\appendix

\section{Gaussian Matrix integrals,  Planar diagrams, and 2D Coulomb gas at $\beta=2$} 
\label{app:review}
Enumeration of maps, i.e. graphs drawn on certain surfaces, random matrices and 2D Coulomb gases are inter-related topics in
mathematical physics. The electrostatic analogy suggested by Wigner~\cite{Wigner57} and Dyson~\cite{Dyson62} in the late 1950's has proved extremely useful in many branches of~\textsf{RMT}. On the other hand, extensive elaborations on the connections between random matrices and the enumeration of maps did not begin until the works in the 1970's by 't~Hooft's~\cite{tHooft74}, Koplin et al.~\cite{Koplin77} and Br\'ezin et al.~\cite{Brezin78} on the planar approximation in various field theories. Once this connection was brought to the surface, matrix models and field theoretical techniques have been exploited to make progress on unsolved graphical enumeration problems~\cite{Bessis80,Itzykson80,Penner88,Itzykson90}. (A comprehensive account can be found in the classical references~\cite{DiFrancesco95,Zvonkin97}.)

Diagrammatic techniques have been successfully applied to a few specific but important ensembles, namely the Gaussian ensembles, see for instance \cite{Verbaarschot84}. The crucial reason is that computing expectation values in {this ensemble} is tantamount to applying Wick's formula in various settings; furthermore,  when pairing the Gaussian entries,  a nice geometrical structure emerges. These patterns have been rediscovered many times by different communities and mastering the relevant vast literature can be a daunting task for the uninitiated.  The aim of this Appendix is twofold. Firstly, we summarize the general ideas connecting matrix integrals and planar enumeration of maps, and at the same time we try to clarify different technical terminologies from various communities. Secondly, we collect results available in the mathematical physics literature and compare them with our findings. Throughout this Appendix, we will concentrate on the \textsf{GUE} ensemble and on the diagrammatic techniques associated to the computation of its moments. For \textsf{GUE}, the joint law of the eigenvalues~\eqref{eq:cov_jpdf}  is the canonical distribution of a 2D Coulomb gas confined in a harmonic potential $V(x)=x^2/4$ at inverse temperature $\beta=2$.

The idea that the large $N$ asymptotic expansion of a matrix integral  around a saddle point can be represented by diagrams
identifiable with certain maps was pioneered by Br\'{e}zin et al.~\cite{Brezin78}. In order to study the distribution of
 $G_{N,\kk}=\tr\G_N^{\kk}$, for fixed $\kk$, it is customary to consider the function
\be
\exp{\{-N^2 E_{N,\kk}(s)\}}=\avg{\exp{\{-N^2s\,G_{N,\kk}\}}} ,\label{eq:logLapl}
\ee
where hereafter the angle brackets denote the average with respect to the \textsf{GUE} measure. From the statistical independence of the Gaussian entries of $\G_N$, it is known~\cite{Brezin78,Itzykson80} that $E_{N,\kk}(s)$ admits a $1/N^2$-expansion 
\be
E_{N,\kk}(s)=\sum_{g\geq0}\frac{1}{N^{2g}}E_{\kk}^{(g)}(s),
\ee
where, for all $g\geq0$, $E_{\kk}^{(g)}(s)$ is analytic in a neighbourhood of $s=0$,
\be
E_{\kk}^{(g)}(s)=-\sum_{v\geq1}\frac{(-s)^v}{v!}D_{\kk}^{(g)}(v).
\ee
In the above expansion, $D_{\kk}^{(g)}(v)$ is the number of inequivalent connected diagrams of genus $g$ with $v$ vertices of valence $\kk$. 

\begin{figure}[t]
\centering
\includegraphics[width=.9\columnwidth]{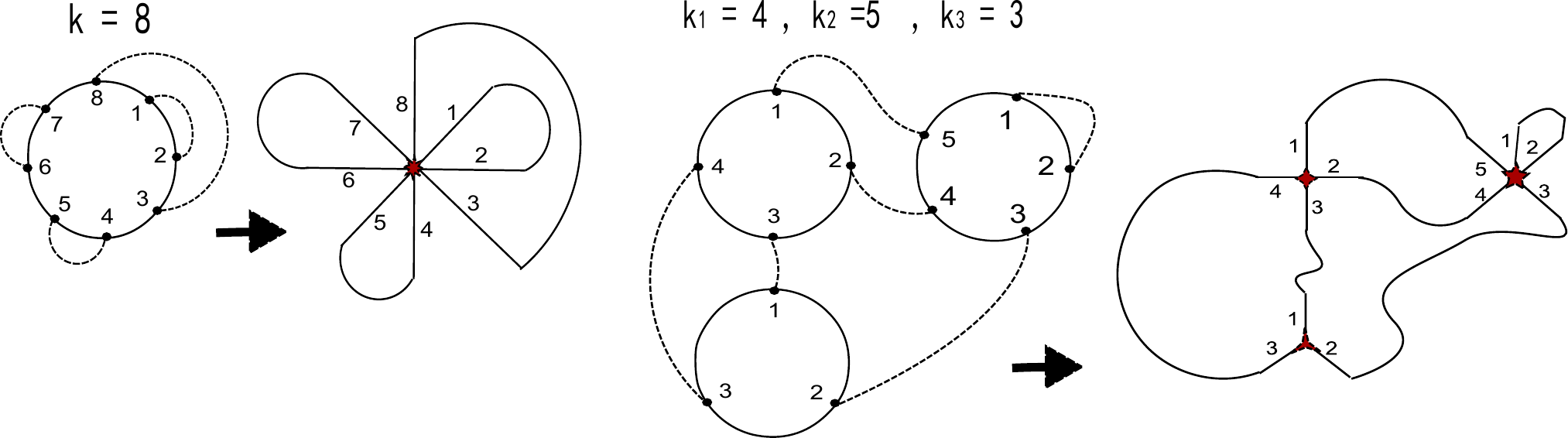}
\caption{{\bf Left:} An example of the mapping between \textsf{NCPP} and planar diagrams with a vertex (or planar maps with a star). In this example, the star has valence $\kk=8$. This diagram arises in a $\varphi^8$-theory. In \textsf{RMT} this diagram contributes to the expectation $\avg{G_{N,8}}$. {\bf Right:} This figure should clarify what happens when several moments are considered. We have a non-crossing connected pairing of three circles with a different number of points on each circle (note that the total number of points is even). The corresponding planar diagram has three vertices ($v=3)$ with different valences. It arises in a mixed field theory with cubic, quartic and quintic interactions. Diagrams of this type are counted when computing the third-order mixed cumulant $C_3(G_{N,4},G_{N,5},G_{N,3})$ of the \textsf{GUE} ensemble.}
\label{fig:diagappendix}
\end{figure}

Therefore, in the large $N$ limit only the $g=0$ contribution survives,
\be
\lim_{N\to\infty}E_{N,\kk}(s)=E_{\kk}^{(0)}(s)=-\sum_{v\geq1}\frac{(-s)^v}{v!}D_{\kk}^{(0)}(v).\label{eq:exp_vertices}
\ee
The quantity  $E_{\kk}^{(0)}(s)$ is often referred to as \emph{cumulant generating function} of the $\kk$-th moment $G_{N,\kk}$.  The coefficients in this expansion count the connected diagrams of a $\varphi^{\kk}$-theory in which all non planar diagrams are omitted: $D_{\kk}^{(0)}(v)$ is the number of connected diagrams with $v$ vertices all of valence $\kk$ that can be drawn on a plane (genus $g=0$)~\footnote{In the mathematical literature, these planar diagrams are also known as \emph{planar maps with $v$ stars} (each star has $\kk$ points).}. The probabilistic meaning of these numbers is clear from~\eqref{eq:logLapl}: if we denote by $C_v(G_{N,\kk},\dots,G_{N,\kk})$ the $v$-th cumulant of $G_{N,\kk}$, then by differentiating \eqref{eq:logLapl} we arrive at
\be
\lim_{N\to\infty}N^{2(v-1)}C_v(\underbrace{G_{N,\kk},\dots,G_{N,\kk}}_{v\,\text{times}})=D_{\kk}^{(0)}(v).
\ee
For instance, the first two cumulants (average and variance) of $G_{N,\kk}$ are given by the number of planar diagrams with one and two $\kk$-valent vertices, respectively,
	\begin{align}
\avg{G_{N,\kk}}&=D_{\kk}^{(0)}(1)+o(1),\label{eq:avg1}\\
N^2\mathrm{Var}({G_{N,\kk}})&=D_{\kk}^{(0)}(2)+o(1).\label{eq:fluct1}
	\end{align}

We emphasize the origin of the two main features of these diagrams: the \emph{connectedness} is intrinsic in the definition of $E_{N,\kk}(s)$ (the logarithm of an exponential generating function);  the \emph{planar} approximation $g=0$ emerges in the large $N$ limit. 
Correction terms are encoded in $E_{\kk}^{(g)}(s)$ for higher genera $g\geq1$, that is by considering $\kk$-valent connected diagrams $D_{\kk}^{(g)}(v)$ on higher genera surfaces  (for instance the first correction comes from legal diagrams $D_{\kk}^{(1)}(v)$ on the torus, $g=1$). In the one-vertex case ($D_{\kk}^{(0)}(1)$ with $\kk\geq1$), these planar diagrams are in bijection with \emph{non-crossing pair partitions} (\textsf{NCPP}) of $\kk$ points. A \textsf{NCPP} of $\kk$ points can be represented by placing the numbers $1,\dots,\kk$ around a single circle and connecting numbers in pairs without crossings. Hence 
\[
D_{\kk}^{(0)}(1)\equiv\#\left\{\text{non-crossing pairings of $\kk$ points on a circles}\right\}.
\]
The theory of non-crossing partitions is an important topic in modern combinatorics and their role in random matrices has been elucidated by \emph{free probability}~\cite{Voiculescu86,Voiculescu91}, with the notion of \emph{free cumulants}~\cite{Speicher93,Nica06}. 

One can extend the above identification to diagrams with more vertices $v\geq1$ and write 
\be
D_{\kk}^{(0)}(v)\equiv\#\{\text{Non-crossing pairings of $v$ circles with $\kk$ points on each circle such that all circles are connected}.\}
\ee 
(See Fig. \ref{fig:diagappendix}.) The connectedness requirement is clear and the non-crossing condition is the translation of the planarity condition for diagrams. Eqs. \eqref{eq:avg1} and \eqref{eq:fluct1} thus connect average and variance of a \emph{single} linear statistics $G_{N,\kappa}$ of the Gaussian ensemble at $\beta=2$ to the problem of non-crossing pairings on one circle, or on two circles, respectively. 

One can generalize the above arguments and consider the joint distribution of $(G_{N,\kk_1},\dots,G_{N,\kk_n})$ for $n\geq2$ with distinct $\kk_i$'s in general. Using the notation $\vec{\kk}=(\kk_1,\dots,\kk_n)$ and $\vec{s}=(s_1,\dots,s_n)$, we may introduce the multidimensional analog of~\eqref{eq:logLapl} 
\be
\exp{\{-N^2E_{N,\vec{\kk}}(\vec{s})\}}=\langle{\exp{\{-N^2\sum_{i=1}^ns_iG_{N,\kk_i}\}}}\rangle. \label{eq:logLapljoint}
\ee
The large $N$ limit of $E_{N,\vec{\kk}}(\vec{s})$ will be the joint cumulant generating function of $(G_{N,\kk_1},\dots,G_{N,\kk_n})$,
\be
E_{\vec{\kk}}^{(0)}(\vec{s})=-\sum_{v\geq1}\sum_{\substack{v_1,\dots,v_n\geq0\\v_1+\cdots+v_n=v}}\left[\prod_{i=1}^n\frac{(-s_i)^{v_i}}{v_i!}\right]D_{\vec{\kk}}^{(0)}(v_1,\dots,v_n)\ ,
\label{eq:field_exp}
\ee
where $D_{\vec{\kk}}^{(0)}(v_1,\dots,v_n)$ is now the number of connected planar diagrams with $v_1$ `$\kk_1$-valent' vertices, $v_2$   `$\kk_2$-valent' vertices, and so on. The relation with the theory of non-crossing partitions is now 
\be
\begin{split}
D_{\vec{\kk}}^{(0)}( v_1,\dots,v_n)& \equiv\#\{\text{Non-crossing pairings of $v=v_1+\cdots+v_n$ circles with $\kk_1$ points on $v_1$ circles,} \\
 & \qquad \; \text{ $\kk_2$ points on $v_2$ circles, \ldots, $\kk_n$ points on $v_n$ circles such that all circles are connected.}\}
 \end{split}
\ee
 Clearly, no such pairing is possible if the total number of points $\sum_{i=1}^{n}v_i\kk_i$ is odd. These numbers provide the leading order of the joint mixed cumulants  $(G_{N,\kk_1},\dots,G_{N,\kk_n})$  of order $v=v_1+\cdots+v_n$,
\be
\lim_{N\to\infty}N^{2(v-1)}C_{v}(\underbrace{G_{N,\kk_1},\dots,G_{N,\kk_1}}_{v_1\,\text{times}},\dots,\underbrace{G_{N,\kk_n},\dots,G_{N,\kk_n}}_{v_n\,\text{times}})=D_{\vec{\kk}}^{(0)}(v_1,\dots,v_n)\ .\label{eq:joint_cumu}
\ee
From~\eqref{eq:joint_cumu}, one can see that for the \textsf{GUE} ensemble the family $\{G_{N,\kk}\}_{\kk\geq0}$ is \emph{asymptotically Gaussian} (the case $\kk=0$ is degenerate) as $N$ goes to infinity. In the language of free probability, this means that the \textsf{GUE} ensemble has a \emph{second order limit distribution} \cite{Mingo04,Mingo06,Mingo07}, i.e. the limits
	\be
	\alpha_{\kk}^{\G}=\lim_{N\to\infty}\avg{G_{N,\kk}}\quad\text{and}\quad (1/2)\alpha_{\kk,\ell}^{\G}=\lim_{N\to\infty}N^2\left[\avg{G_{N,\kk}G_{N,\ell}}-\avg{G_{N,\kk}}\avg{G_{N,\ell}}\right]
	\ee
exist for all $\kk,\ell\geq0$, and the higher-order cumulants decay faster to zero for large $N$.

Br\'ezin et al.~\cite{Brezin78} considered the planar approximation for $\kk=3$ and $\kk=4$ (quartic and cubic vertices). They computed exactly the cumulant generating functions $E_{3}^{(0)}(s)$ and $E_{4}^{(0)}(s)$ thus providing the \emph{full} family of cumulant of $\Tr\G^{3}$ and $\Tr\G^{4}$ to leading order in $N$. For the quartic interaction, the higher genus corrections $E_{4}^{(1)}(s)$ and $E_{4}^{(2)}(s)$ have been computed explicitly in~\cite{Bessis80}. The problem of mixed interaction (diagrams whose vertices have mixed valence) has been less explored even in the planar regime $g=0$ --- the case of the combined cubic and quartic interactions, i.e. the computation of $E_{3,4}^{(0)}(s_1,s_2)$ was proposed in~\cite{Brezin78} but was not worked out explicitly. Later, Harer and Zagier~\cite{Harer} computed the exact finite $N$ value of $\avg{G_{N,\kk}}$ (i.e. the linear term in $s$ of $E_{N,\kk}(s)$ in~\eqref{eq:logLapl} for all $\kk$).

\begin{figure}[t]
\centering
\includegraphics[width=.12\columnwidth]{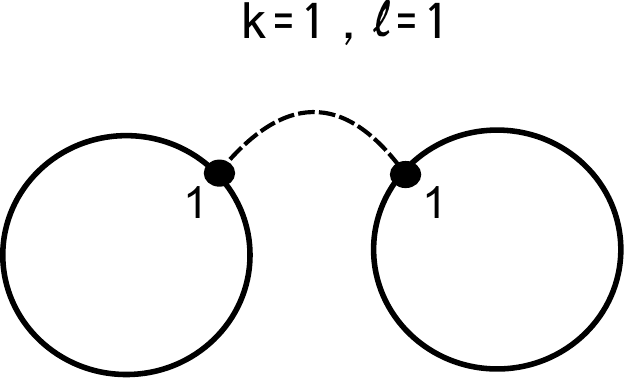}\quad\quad
\includegraphics[width=.12\columnwidth]{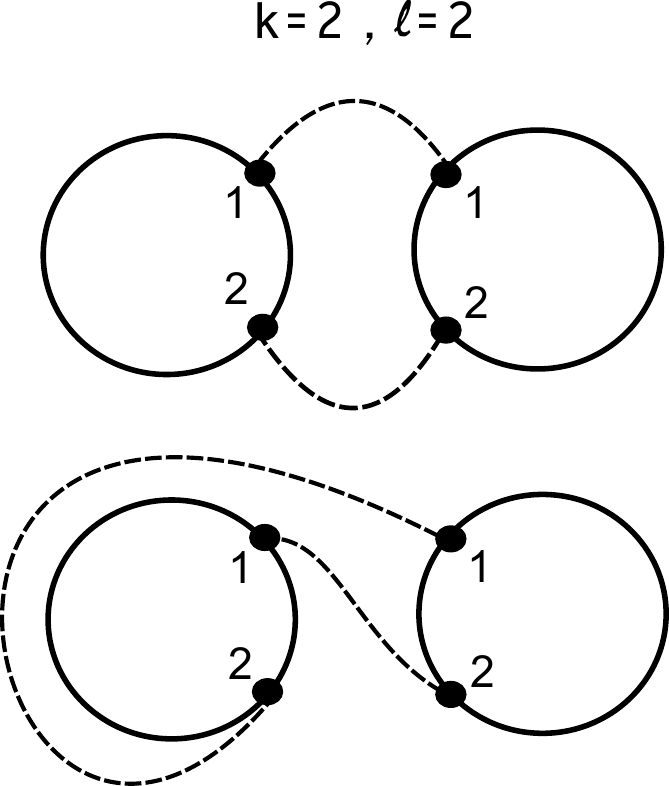}\quad\quad
\includegraphics[width=.12\columnwidth]{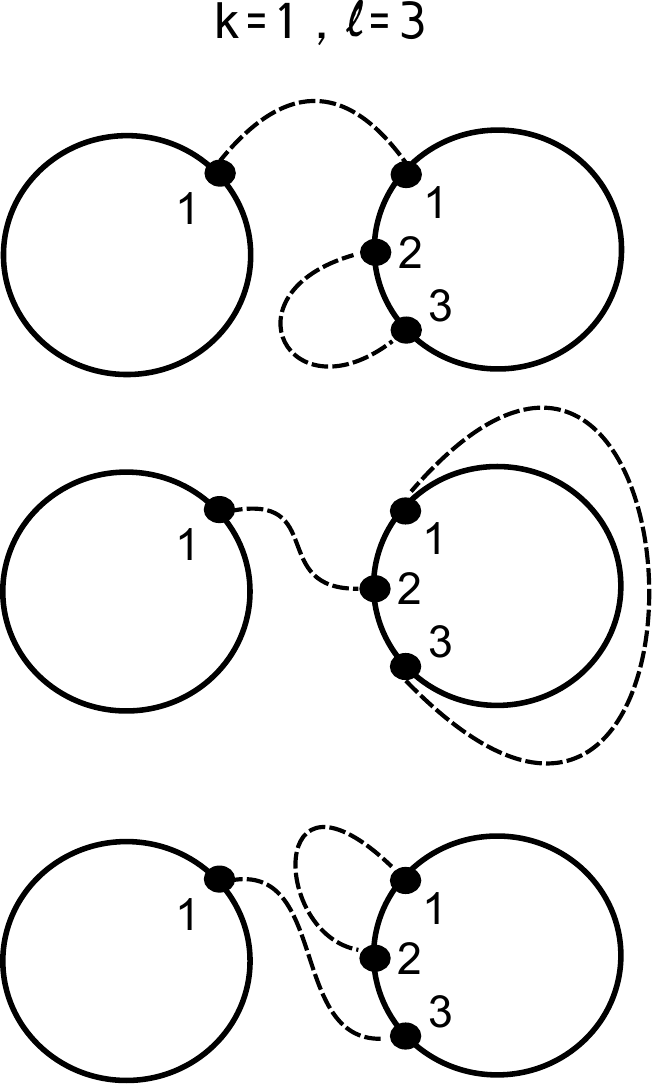}\quad\quad
\includegraphics[width=.53\columnwidth]{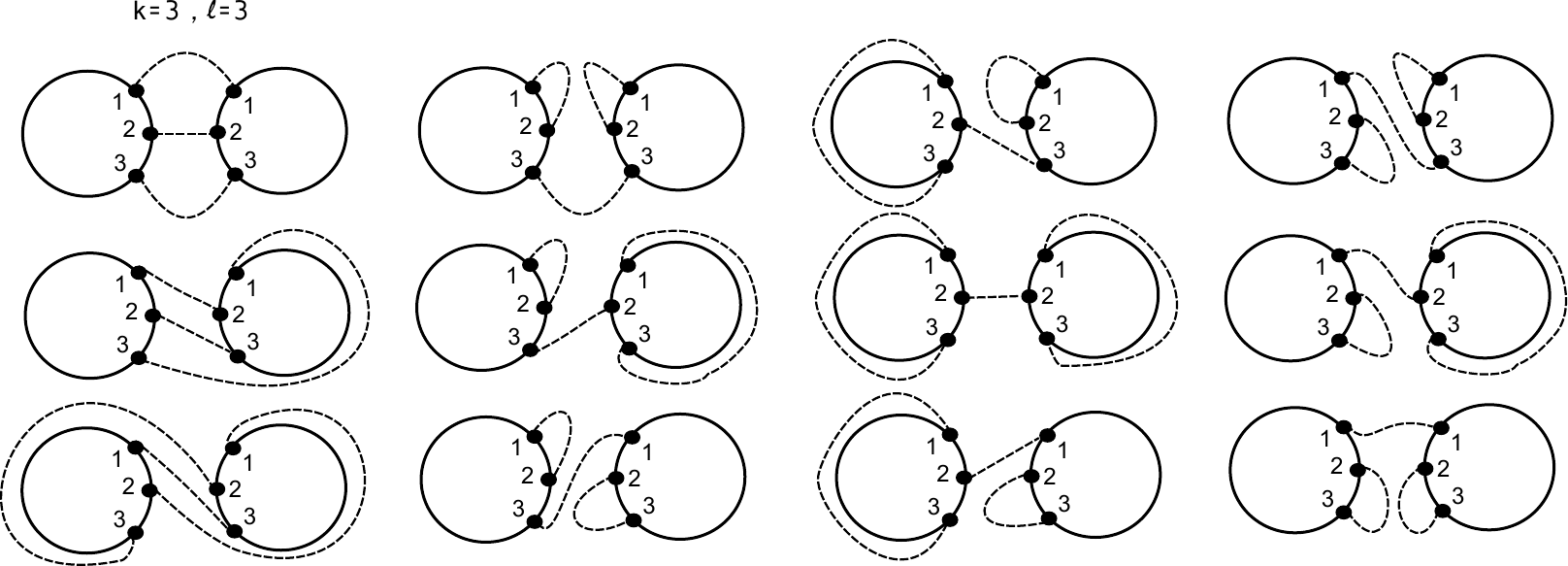}
\caption{Non-crossing pairings of two circles with $\kk$ points on the first circle and $\ell$ points on the second circle, such that the two circles are connected. The number of such pairings is equal to $D_{\kk,\ell}^{(0)}(1,1)=(1/2)\alpha_{\kk,\ell}^{\G}$, i.e. $\lim_{N\to\infty}\Cov(\Tr\G_N^{\kk},\Tr\G_N^{\ell})$ with $\G_N$ belonging to the \textsf{GUE}. We have $(1/2)\alpha_{1,1}=1$, $(1/2)\alpha_{2,2}=2$, $(1/2)\alpha_{1,3}=3$, $(1/2)\alpha_{3,3}=12$ according to Eq. \eqref{eq:covG} and Table \ref{tab:num}. }
\label{fig:diagrams}
\end{figure}

Our results (when restricted to the \textsf{GUE} ensemble) provide a complete answer to the following problem: for \emph{any} finite family of power traces $(G_{N,\kk_1},\dots,G_{N,\kk_n})$ compute the joint cumulant generating function $E_{\vec{\kk}}^{(0)}(\vec{s})$ in~\eqref{eq:field_exp} up to quadratic terms,
\begin{align}
E_{\vec{\kk}}^{(0)}(\vec{s})&=-\sum_{1\leq v\leq2}\sum_{\substack{v_1,\dots,v_n\geq0\\v_1+\cdots+v_n=v}}\left[\prod_{i=1}^n\frac{(-s_i)^{v_i}}{v_i!}\right]D_{\vec{\kk}}^{(0)}(v_1,\dots,v_n)+ \dotsb\\
&=\sum_{i=1}^ns_iD_{\kk_i}^{(0)}(1)
-\frac{1}{2}\sum_{\substack{i,j=1}}^ns_is_j{D^{(0)}_{\kk_i,\kk_j}}(1,1)+ \dotsb\label{intint}
\end{align}
This corresponds to considering the contribution from \emph{all} types of planar diagrams (every kind of combined interactions) with \emph{at most} two vertices (of arbitrary valence). Here $D_{\kk_i}^{(0)}(1)$ and ${D^{(0)}_{\kk_i,\kk_j}}(1,1)$ give the leading order behaviour of the averages and covariances of the family $(G_{N,\kk_1},\dots,G_{N,\kk_n})$.  The results of Section~\ref{sub:Gaussian}, restricted to $\beta=2$, give
	\begin{align}
	\lim_{N\to\infty}\avg{G_{N,\kk}}&=D_{\kk}^{(0)}(1)=\alpha_{\kk}^{\G},\\
  \lim_{N\to\infty}N^2\Cov(G_{N,\kk}, G_{N,\ell})&={D^{(0)}_{\kk,\ell}}(1,1)=(1/2)\alpha_{\kk,\ell}^{\G}.\label{jointb}
	\end{align}
The explicit formulae were given in~\eqref{eq:alphak} and~\eqref{eq:covG}, respectively.
Therefore, from the above discussion we have
 \begin{align}
 \alpha_{\kk}^{\G}& =\#\{\text{Non-crossing pairings of $\kk$ points on the circle}\},\\
 (1/2)\alpha_{\kk,\ell}^{\G}& =\#\{\text{Non-crossing pairings of two circles with  $\kk$ points on the $1$st circle}\nonumber \\ & \qquad \; \text{and $\ell$ points on the $2$nd circle such that the two circles are connected}\}.
\end{align}
In Fig.~\ref{fig:diagrams} we report few such diagrams.

The difference of our method compared to previous works on the planar approximation in field theories is the following. Instead of expanding the cumulant generating function of a fixed $G_{N,\kk}$ (or a fixed $n$-uple $(G_{N,\kk_1},\dots,G_{N,\kk_n})$) in the number of vertices $v$ as in~\eqref{eq:exp_vertices}-\eqref{eq:field_exp}, in this paper we have considered the whole family of generating functions of a generic $n$-uple $(G_{N,\kk_1},\dots,G_{N,\kk_n})$ up to quadratic terms. In diagrammatic terms, instead of an expansion of planar diagrams with fixed valence $\kk$ in the number of vertices $v$, we would obtain a series in the valences of vertices $\kk$ and $\ell$ of planar diagrams with at most two vertices $v\leq2$. 

We mention that the combinatorial problem associated to $\alpha_{\kk}^{\G}$ and $(1/2)\alpha_{\kk,\ell}^{\G}$ was solved by Tutte~\cite{Tutte62} as a particular case of a more general problem on the number of so-called \emph{slicings} of a \emph{band.} 
We note that Tutte's formula enumerates general slicings, with a strong restriction on the parity of the vertices. In the language of \textsf{RMT}, Tutte provided a formula for the generic mixed cumulant  $D_{\vec{\kk}}^{(0)}(v_1,\dots,v_n)$ of $(G_{N,\kk_1},\dots,G_{N,\kk_n})$ (to leading order in $N$), with the condition that $\kk_1,\dots,\kk_N$ should all be even~\footnote{In the case of only two bounding curves (our case), Tutte managed to obtain a formula that comprises the cases $\kk$ and $\ell$ both even or both odd. No orientation is involved in his problem, hence in his formula there is a factor of $1/2$.}. We find interesting to note that this obstruction (that might appear unjustified in the combinatorial problem) can be explained in field-theory with the instability of any $\varphi^{\kk}$-interaction for $\kk$ odd.

\section{Outline of the proof of (\ref{eq:covG})} 
\label{appB}
It is convenient to cast $\alpha_{\kappa,\ell}^{\mathcal{G}}$ in \eqref{eq:covG} for $\kk,\ell>0$ in term of Euler Gamma functions
\begin{equation}
\alpha_{\kappa,\ell}^{\mathcal{G}}=
\begin{cases}
\displaystyle\frac{16 \Gamma (\kappa) \Gamma (\ell)}{(\kappa+\ell) \Gamma
   \left(\frac{\kappa}{2}\right)^2 \Gamma
   \left(\frac{\ell}{2}\right)^2} &\mbox{ for }\ell\text{ and }\kappa\mbox{ even }\\
\displaystyle\frac{\kappa \ell 4^{\kappa+\ell-1} \Gamma \left(\frac{\kappa}{2}\right)^2
   \Gamma \left(\frac{\ell}{2}\right)^2}{\pi ^2 (\kappa+\ell)
   \Gamma (\kappa) \Gamma (\ell)}&\mbox{ for }\ell\text{ and }\kappa\mbox{ odd }\ .\\
\end{cases}
\end{equation}

Setting $\kappa=2 r$ and $\ell=2s$ (in the even-even case) and $\kappa=2 r+1$ and $\ell=2s+1$ (in the odd-odd case), we can evaluate the double sum in \eqref{eq:main_res_gen} as

\begin{equation}
\sum_{\kappa,\ell=1}^\infty \alpha_{\kappa,\ell}^{\mathcal{G}} z^\kappa\zeta^\ell=\sum_{r,s=1}^\infty \frac{8 \Gamma (2r) \Gamma (2s)}{(r+s) \Gamma
   \left(r\right)^2 \Gamma
   \left(s\right)^2}z^{2r}\zeta^{2s}+
   \sum_{r,s=0}^\infty \frac{(2r+1) (2s+1) 4^{2r+2s+1} \Gamma \left(\frac{2r+1}{2}\right)^2
   \Gamma \left(\frac{2s+1}{2}\right)^2}{\pi ^2 (2r+2s+2)
   \Gamma (2r+1) \Gamma (2s+1)}z^{2r+1}\zeta^{2s+1}\ .
 \end{equation}
 The two series on the r.h.s. can be computed in closed form using the following identities
 \begin{align}
 \frac{1}{r+s} &=\int_0^\infty\de t \,\mathrm{e}^{-t(r+s)}\ ,\\
 \sum_{r=1}^\infty  \frac{ \Gamma (2r) }{\Gamma
   \left(r\right)^2 }(z\mathrm{e}^{-t/2})^{2r} &=\frac{ \mathrm{e}^{-t} z^2}{\left(\mathrm{e}^{-t} \left(\mathrm{e}^t-4 z^2\right)\right)^{3/2}}\ ,\\
    \sum_{r=0}^\infty \frac{(2r+1)  \Gamma \left(\frac{2r+1}{2}\right)^2
  }{
   \Gamma (2r+1)}(4 z \mathrm{e}^{-t})^{2r} &=\frac{\pi\mathrm{e}^{2 t}}{ \left(\mathrm{e}^{2 t}-4 z^2\right) \sqrt{1-4 \mathrm{e}^{-2 t} z^2}}\ .
 \end{align}

Therefore we can rewrite the double sum in integral form
\begin{align}
\nonumber \sum_{\kappa,\ell=1}^\infty \alpha_{\kappa,\ell}^{\mathcal{G}} z^\kappa\zeta^\ell &=8\int_0^\infty\de t \frac{ \mathrm{e}^{-t} z^2}{\left(\mathrm{e}^{-t} \left(\mathrm{e}^t-4 z^2\right)\right)^{3/2}}\frac{ \mathrm{e}^{-t} \zeta^2}{\left(\mathrm{e}^{-t} \left(\mathrm{e}^t-4 \zeta^2\right)\right)^{3/2}}\\
&+\frac{4}{\pi^2}z\zeta\int_0^\infty\de t\ \mathrm{e}^{-2t} \frac{\pi\mathrm{e}^{2 t}}{ \left(\mathrm{e}^{2 t}-4 z^2\right) \sqrt{1-4 \mathrm{e}^{-2 t} z^2}}
\frac{\pi\mathrm{e}^{2 t}}{ \left(\mathrm{e}^{2 t}-4 \zeta^2\right) \sqrt{1-4 \mathrm{e}^{-2 t} \zeta^2}}\ .
\end{align}
The integrals can be carried out in closed form, and after simplifications we obtain
\begin{equation}
\sum_{\kappa,\ell=1}^\infty \alpha_{\kappa,\ell}^{\mathcal{G}} z^\kappa\zeta^\ell =-\frac{\zeta  z \left(\sqrt{1-4 \zeta ^2} \sqrt{1-4 z^2}+4 \zeta  z-1\right)}{\sqrt{1-4
   \zeta ^2} \sqrt{1-4 z^2} (z-\zeta )^2},\qquad\mbox{ for }-1/2<z,\zeta<1/2\ ,
\end{equation}
which (after multiplying by $1/\beta$) coincides with $F_{\pm 2}(z,\zeta)$ as given in \eqref{eq:main_res_G}.

\end{document}